\documentstyle[a4,11pt,epsfig,twoside]{article}
\begin{document}
\thispagestyle{empty}
\mbox{  }\hfill TUM--HEP--493/02\\
\mbox{  }\hfill hep--ph/0211407\\
\begin{center}
{\Large\bf CP-ODD PHASES @ NLC: A PROJECT OUTLINE }\\[8mm]
B.GAISSMAIER \footnote{ bgaissma@ph.tum.de} \\[4mm]
{\it Physik Dept., TU M\"{u}nchen, James Franck Str., D--85748
Garching, Germany} \\

\end{center}
\bigskip
\bigskip
\bigskip
\begin{abstract}
  Low-energy results from measurements of leptonic dipole operators
  are used to derive constraints on phases of the MSSM.  After
  rediscovering older bounds on these phases, we try to investigate
  the impact of these possibly non-vanishing CP-odd phases on the
  measurement of CP-even cross-sections at the next leptonic collider.
\end{abstract}

\section{Introduction}
The most general MSSM provides a rather big number of complex phases,
most of which originate from the soft breaking terms. Of course, the
possible ranges of these phases are restricted by the current
measurements of leptonic dipole operators such as $d_e$
\cite{Groom:in} or $a_\mu$
\cite{Bennett:2002jb},\cite{Hayakawa:2001bb}. Fortunately, theory can
provide at least one scenario (cancellation), in which the bounds for
the dipole operators are respected without too small values for the
phases \cite{Brhlik:1998zn},\cite{Ibrahim:1998je}. Once assumptions
about the mechanism of supersymmetry breaking are made, most soft
breaking parameters can be computed from their input parameter using
the RGE's for running them from the unification scale to the scale
relevant at colliders, and the bounds on the phases can
get rather strong.\\
Nevertheless, since mass-spectra and couplings obviously depend on -
besides several real parameters - the phases, they are a priori
non-neglible in studies of possible signals at future colliders and
should be considered as free parameters of the model. Actually the
situation is even worse: neglecting phases during the determination of
real parameters from experimental data could lead to wrong inputs for
theorists, who want to specify the underlying theory at the
unification scale.\\
Despite the necessity to extract values for phases from experiment,
the construction of sizable and comfortably accessible CP-violating
observables is rather difficult in most of the production channels at
an $e^+e^-$-machine as at least one secondary decay has to be
included. Considering this problem we are introducing an object which
quantifies the impact of non-vanishing phases on CP-even
cross-sections and allows us to investigate this impact to some
extent.
\section{Idea, its Complementarity and Framework}
The basic idea of this work \cite{us} is to take today's low energy
data (lower mass-bounds, $d_e$ and $a_\mu$) as a set of constraints
for a parameter space scan and then to apply the resulting, low-energy
compatible points to a set of high energy experiments @ NLC ($e^+e^-$
and $e^-e^-$-option). This set of high energy-experiments is given by
the total, unpolarized cross-sections for $\tilde{\chi}^0$-,
$\tilde{\chi}^-$-, $\tilde{e}^-\tilde{e}^+$ pair production and
is completed by $\tilde{e}^-\tilde{e}^-$ production \cite{Choi:2001ww},\cite{Choi:2000ta},\cite{Peskin:1998jy},\cite{Thomas:1997ng}.\\
Since we are are dominantly interested in the role of phases, the most
stringent bounds on parameter space arise from the possible size of
the SUSY-contributions to $d_e$ and $a_\mu$, which are given by:
$$
\begin{array}{c}
5.7\times10^{-11}<(a_\mu)_{SUSY}<49.3\times 10^{-11}\\
-13\times10^{-28}{\rm ecm}<(d_e)_{SUSY}<59\times10^{-28}{\rm ecm}
\end{array}
$$
Recalling the diagrammatic structure of the SUSY-contribution to
the leptonic dipol operators ($\tilde{\chi}^0\tilde{e}$ and
$\tilde{\chi}^-\tilde{\nu}$) and comparing with the several
(tree-level) diagrams for the given production modes we notice a {\em
  complementarity} between low- and high-energy observables: both
low-energy observables always depend on products of different vertices
while each diagram contributing to the cross-sections depends only on
bilinears of one supersymmetric vertex (or a product of a
supersymmetric one and a known SM-vertex). This implies that the
low-energy observables connect different parts of SUSY-Lagrangians and
therefore only can give bounds on combinations of phases, whereas
high-energy observables
can be used to investigate the different parts separately.\\
As framework for this project we are using the MSSM with R-parity,
neglect sflavour mixing and assume horizontal universality of the
Yukawa-like soft breaking $A$ terms. Therefore the real parameters
relevant for our analysis are:
$$
|\mu|,\;\tilde{m}_L,\;\tilde{m}_R,\;|M_1|,\;M_2,\;|A|,\;\tan\beta,
$$
and the investigated phases are ($\phi_2=0$ by convention):
$$
\phi_\mu,\;\phi_1,\;\phi_A.
$$
\section{Significance and Mass Variation}
As pointed out already, our aim is to find an object quantifying the
impact of CP-odd phases on CP-even cross-sections and to understand
the behaviour and meaning of this object. As underlying assumption for
the definition of this object we assume that real parameters are
already fixed. This assumption is indeed simplifying our analyses
significantly, nevertherless we think it is a fair one, as most of the
recent analysis do the same {\em and} neglect phases.  The basic idea
for the object we are introducing as {\em significance of a CP-even
  cross section with respect to CP-phases} is to compare the deviation
of counting rates between a CP-conserving (CPC: real parameters, all
phases $\equiv 0$) and CP-violating point (CPV: same real parameters,
{\em but} phases $\neq 0$ and low-energy compatible) to the
statistical error of the CPC-point. In formulae this idea reads as:
$$
S\propto \frac{\Delta N_{CPV-CPC}}{\delta N_{CPC}}.
$$
As there are two CP-conserving values (0, $\pi$) for each phase, we
have to deal with 8 CPC points for each set of real parameters,
 a priori the same number of
(different) significances is available for each cross-section and we have
to decide which one to use in further discussions.\\
Our selection rule is then take the minimum of S with respect to these
8 points, this corresponds to the most conservative estimate of the
impact of phases on cross-sections, and $S$ finally reads:
$$
S(\sigma_{f_if_j})={\rm
  min}\left(\frac{|\sigma_{f_if_j}^{CPV}-\sigma_{f_if_j}^{CPC}|}{\sqrt{\sigma_{f_if_j}^{CPC}}}\right)\sqrt{{\cal L}},
$$
where the indices $f_i$, $f_j$ refer a given out-mode with
particles $f_i$, $f_j$ and $\sigma_{f_if_j}^{CPV/CPC}$ are the
corresponding cross-sections in a CPV- or CPC-point.\\
Although it is rather clear that a high value of $S(\sigma_{f_if_j})$
indicates a significant dependence of the mode on the phases, it is
not clear where this dependence originates from. So the next step is
to separate kinematical from coupling effects. For this purpose we
define a variation of masses as:
$$
V(m_i,m_j)=\frac{(m_i+m_j)^{CPV}-(m_i+m_j)^{CPC}}{(m_i+m_j)^{CPC}},
$$
which is calculated in the CPV-point which minimizes
$S(\sigma_{f_if_j})$. Since slepton masses are phase independent we
are investigating slepton production with respect to the lightest Neutralino
in the $t$-channel diagrams. The point about studying this
mass-variation is that a small variation of masses
correspondes to small variation of the kinematical functions
($\beta(\frac{m_i^2}{s}), \lambda(\frac{m_i^2}{s},\frac{m_j^2}{s})$)
and a large $S(\sigma_{f_if_j})$ should therefore be induced by
coupling effects.
\section{Numerical Results}
\begin{figure}[h!]
  \epsfxsize=7cm\epsffile{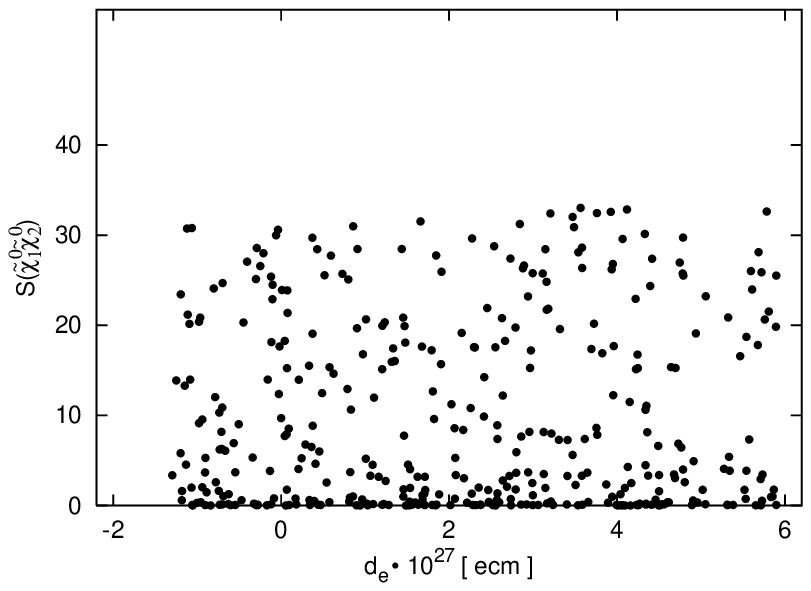}\hspace*{\fill}%
  \epsfxsize=7cm\epsffile{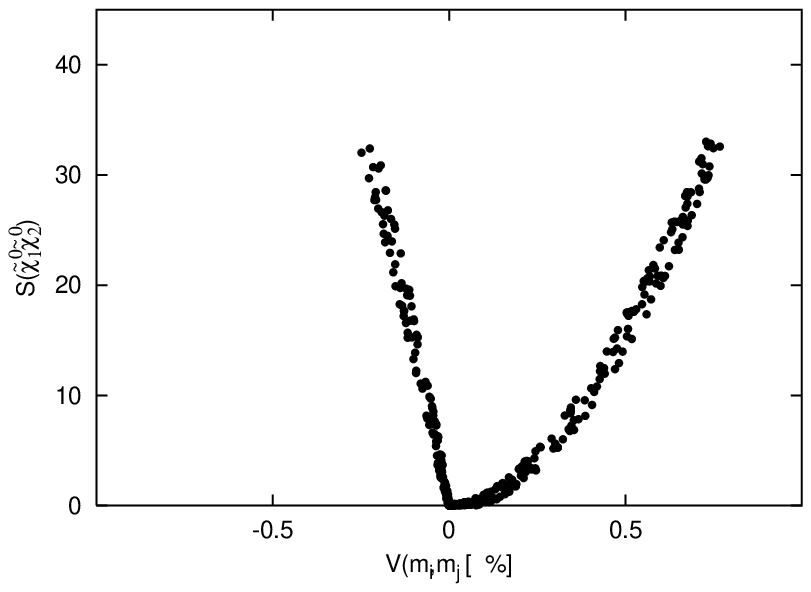}\vspace{0.5cm}
\epsfxsize=7cm\epsffile{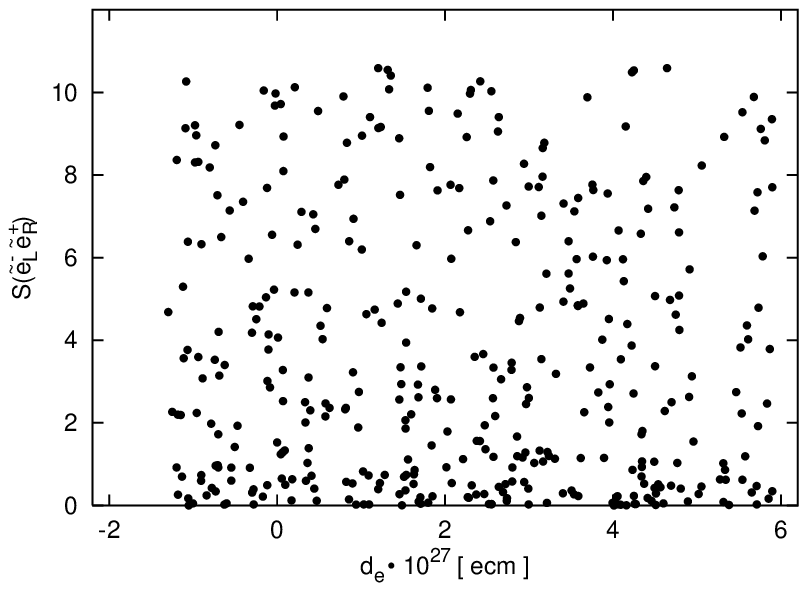}\hspace*{\fill}%
  \epsfxsize=7cm\epsffile{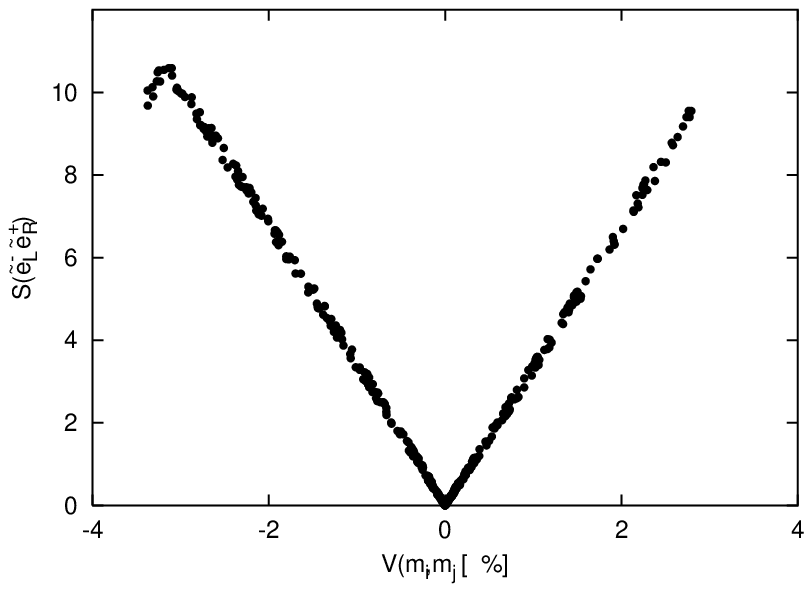}
\caption{upper boxes: $S(\tilde{\chi}_1^0\tilde{\chi}_2^0)$ plotted
  againgst $d_e$(left panel) and $V(m_i,m_j)$ (right panel) for
  $|\mu|$=200 GeV and $\tan\beta$=12, lower boxes: same for
  $S(\tilde{e}_L^-\tilde{e}_R^+)$; $ \sqrt{s}$=500 GeV and ${\cal
    L}$=500 ${\rm fb}^{-1}$.}
\label{fig1}
\end{figure}

\begin{figure}[h!]
\begin{center} 
\epsfxsize=7cm\epsffile{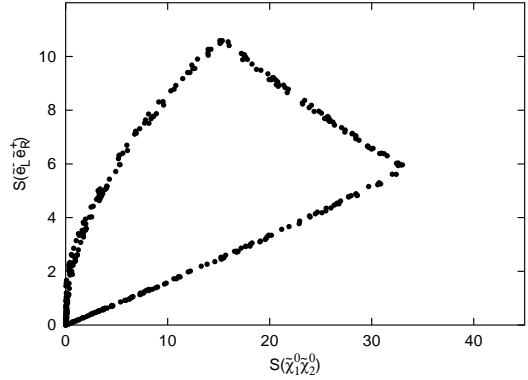}
\end{center}
\caption{Correlation between $S(\tilde{e}_L^-\tilde{e}_R^+)$ and
  $S(\tilde{\chi}_1^0\tilde{\chi}_2^0)$ for $|\mu|$=200GeV and
  $\tan\beta$=12 for $\sqrt{s}$=500 GeV and ${\cal
    L}$=500 ${\rm fb}^{-1}$.}
\label{fig2}
\end{figure}

Our complete numerical analysis is currently performed for different
values of $\tan\beta$ (3,6,9,12) and two values for $|\mu|$ (200 GeV,
500 GeV). The remaining real parameters are fixed as: $M_2$= 200 GeV,
$M_1$=100 GeV, $\tilde{m}_L$=235 GeV, $\tilde{m}_R$=180 GeV and
$|A|$=500 GeV \cite{us}. For each of these 8 real parameter points we scan the
phases $\Phi_\mu$, $\Phi_1$, $\Phi_A$ randomly in (0, 2$\pi$), after
applying the low-energy constraints as cuts the typical survival rate
is a few permille (initially 500,000 points).\\
By correlating low-energy compatible values of the phases, we always
find rather small bands for $\Phi_\mu$ (${\cal O}(\pi/6)$) located
around 0, while the bands for $\Phi_1$ and $\Phi_A$ can be rather
large or even complete. The concrete pattern of correlations between
them of course depends sensitively on the choice of real parameters.
Anyway, these results are no news, so we don't present any of the low-energy correlations here.\cite{Brhlik:1998zn}\\
Although the number of correlations between low- and high-observables
or between high-energy-observables is large, we can illustrate the
basic results using two examples and a set of real parameters. As
examples we show the correlations between $d_e$ and the significance
and the correlation between mass-variation and significance for the
$\tilde{\chi}_1^0\tilde{\chi}_2^0$-mode (left and right, upper panels
in Fig.\ref{fig1}) and the $\tilde{e}_L^-\tilde{e}^+_R$-mode (left and
right, lower panels
in Fig.\ref{fig1}) in $e^-e^+$-collisions.\\
Both modes illustrate that sizable significances are possible over the
complete range of $d_e$, indicating possible big deviations between
CPC and CPV-points even for vanishing $d_e$. The obvious
no-correlation pattern between significances and $d_e$ is rather
transparent as the significances are functions of 2 phases while $d_e$
depends on 3 phases. Note that $d_e$ violates chirality and therefore
always includes a factor of the electron Yukawa coupling, so that
contributions proportional to $A_e$ are {\em not} suppressed here.
Basically the same argumentation applied to the strong correlation
patterns between mass-variations and significances, as masses and
cross-sections both depend on the same phases. Since the scale of
mass-variation is of {\cal O}(\%) for both modes the significance must
originate dominantly from
coupling effects.\\
In Fig.\ref{fig2} we show the correlation between the significances of
both modes, again we observe a correlation pattern as both
cross-sections depend on the same phases. The triangle structure of
the correlation is clear, as the significance of each mode reaches a
maximum for a certain value of $\Phi_1$ ($\Phi_\mu$ is a small band
and can be considered as fixed) and the maximizing values for
different cross-sections do {\em not} coincide.
\section{Conclusion}
After using low-energy data as constraints on parameter space we
rediscover older bounds on phases, the allowed range for the phases
can be rather big. We can give a quantity ({\em significance}) to
estimate the impact of CP-odd phases on CP-even cross-sections. This
significance can be rather big, implying a significant deviation of
CP-violating scenarios from CP-conserving one already at the level of
cross-sections. These significances are correlated with each other,
{\em but not} with $d_e$ and $a_\mu$ and they are dominantly due to
coupling effects. Therefore phases should be taken into account in any
discussion of how to extract model parameters from measured
cross-sections.
\section*{Acknowledgements}
The project is the result of a collaboration with S.Y.Choi and M.Drees.
The work of BG is supported in part by the Deutsche
Forschungsgemeinschaft, project number DR 263.

\end{document}